\documentclass[5p,authoryear]{elsarticle}

\usepackage{graphics}
\usepackage{graphicx}
\usepackage{epsfig}
\usepackage{rotating}
\usepackage{amssymb}

\usepackage{lineno}

\journal{New Astronomy Review}

\begin{document}

\begin{frontmatter}



\title{Long term variability of the Broad Emission Line profiles in AGN}


\author[sao]{A.I. Shapovalova}
\ead{ashap@sao.ru}
\address[sao]{Special Astrophysical Observatory of the Russian AS, Nizhnij Arkhyz, Karachaevo-Cherkesia 369167, Russia}
\author[aob,hum]{L.\v C. Popovi\'c}
\address[aob]{Astronomical Observatory, Volgina 7, 11060 Belgrade 74,
Serbia}
\author[sai]{N.G. Bochkarev}
\address[sai]{Sternberg Astronomical Institute, Moscow, Russia}
\author[sao] {A. N. Burenkov}
\author[ina]{V.H. Chavushyan,\fnref{fn6}}
\address[ina]{Instituto Nacional de Astrof\'{\i}sica, \'{O}ptica y
Electr\'onica, Apartado Postal 51, CP 72000, Puebla, Pue.
M\'exico}
\author[luth]{S. Collin}
\address[luth]{LUTH, Observatoire de Paris, CNRS, Universite Paris
Diderot; 5 Place Jules Janssen, 92190}
\author[sai] {V. T. Doroshenko}
\author[doa]{D.Ili\'c}
\address[doa]{Department of Astronomy, Faculty of Mathematics, University of Belgrade, Studentski trg 16, 11000 Belgrade,
Serbia}
\author[doa]{A. Kova\v cevi\'c}
\begin{abstract}

Results of a long-term monitoring ($\gtrsim 10$ years) of the broad
line and continuum fluxes of three Active Galactic Nuclei (AGN), 3C
390.3, NGC 4151, and NGC 5548, are presented. We analyze the
H$\alpha$ and H$\beta$ profile variations during the monitoring
period and study different details (as bumps, absorption bands)
which can indicate structural changes in the Broad Line Region
(BLR). The BLR dimensions are estimated using the time lags  between
the continuum and the broad lines flux variations. We find that in
the case of 3C 390.3 and NGC 5548 a disk geometry can explain both
the broad line profiles and their flux variations, while the BLR of
NGC 4151 seems  more complex and is probably composed of two or
three kinematically different regions.

\end{abstract}


\end{frontmatter}


\section{INTRODUCTION}

An important question in the study of Active Galactic Nuclei (AGN)
is the nature of the ''central engine''. Nowadays, it is widely
accepted  that the nuclear activity is caused by accretion of gas
onto a supermassive black hole (SMBH). More than 40 years ago,
\cite{W68a} classified AGN into two types: AGN1 whose spectra show
broad (FWHM$> 1000 \,{\rm km/s}$) and narrow (FWHM$\sim 400\, {\rm
km/s}$) emission lines; and AGN2 with only narrow emission lines.
Later, the classification was  improved by including intermediate
types from AGN 1.2 to AGN 1.9 \citep[see][]{Os89}. It was first
believed that AGN1 and AGN2 have different physical natures.
However, more than 20 years ago  a unified scheme of the AGN
structure was proposed, where an accretion disk (AD) around a SMBH
are assumed to be  in the center of all AGN. Around the very central
part,  two emitting regions, the ''Broad Line Region (BLR)'' and the
''Narrow Line Region (NLR)'', are surrounded by a torus of dust
\citep{AM85,An93}. Then the difference in spectra between AGN1 and
AGN2 is related to the viewing angle  of an observer with respect to
the torus orientation.

The BLR, i.e. the Broad Emission Lines (BELs) forming region,
consists of gas obviously linked with the accretion process onto a
SMBH.  It is important to investigate its structure in order to use
the BELs  for the estimation of mass and the accretion rate of the
SMBH. The typical angular size of the BLR is $R_{\rm BLR}< 0.001\,
{\rm arcsec}$, consequently even for nearby AGN1 this region is
presently not resolved  with the most sensitive optical
interferometers. The only ways to study  the BLR are indirect
methods, as e.g. ``reverberation mapping'' (Peterson et al. 1993).

It is well known that AGN vary in luminosity on time scales from
years to hours, over the whole wavelength range from the radio to
X-rays or  $\gamma$-rays. In particular, the flux in BELs varies
with respect to  the ionizing continuum flux  with short time delays
(days to weeks for Seyfert galaxies), due to light-travel time
effects. If the BLR presents systematic motions such as infalling,
outflowing, circular motions, the profiles of the broad emission
lines should vary in a way related with its geometry and kinematics.
The motions also depend on the processes of gas relaxation that
follows changes in the ionizing flux.

An important progress in understanding the BLR structure was
achieved as a result of multiwavelength monitoring  campaigns
performed within the framework of the International AGN Watch, a
consortium organized to study several Seyfert galaxies \citep{Pe99}.

Its main results  \citep[see][and references therein]{Pe08}, and
references therein) can be summarized as follows (i) The response
time of the H$\beta$ line to continuum variations varies from year
to year and is correlated with the average continuum flux; (ii) the
BLR sizes range from light-days to several light-weeks; (iii) the
high ionization lines have a shorter response time than the low
ionization lines, indicating the presence of an ionization
stratification along the BLR; (iv)  rotation or chaotic motions
dominate onto radial motions in the BLR; (v) the optical and
ultraviolet continua vary almost simultaneously (i.e. without
showing a significant delay).

All these conclusions improved our knowledge of the BLR, but many
questions remain still open, as e.g. the existence of a small
fraction of  double peaked profiles which could indicate the
presence an accretion disk, or that of an inflow or an outflow. To
answer these questions, a long term monitoring of AGN with a
detailed study of the different broad line profiles was required.

In this paper we give an overview of our recent investigation based
on the long-time monitoring of three AGNs with different BEL
profiles: 3C 390.3 with double peaked profiles, NGC 4151 with
strongly variable BEL profiles, and NGC 5548 with shoulders in the
profiles.

\section{OBSERVATIONS}

High quality spectra ($S \slash N >  50$) in the continuum around
H${\alpha}$ and H${\beta}$ were obtained in the spectral range 4000
to 7500 \AA, using a long--slit spectrograph equipped with CCDs,
with a resolution between 5 and 15 \AA, using the 6-m and 1-m SAO's
telescopes (Russia) and the GHAO's 2.1-m telescope (Cananea,
Mexico).  Spectrophotometric standard stars were observed every
nights.  The image reduction process included bias subtraction,
flat-field corrections, cosmic ray removal, 2D wavelength
linearization, sky spectrum subtraction, addition of the spectra for
each night, and  relative flux calibration based on standard star
observations.

For the absolute calibration, the fluxes of the narrow emission
lines are adopted  for scaling the AGN spectra, because they are
known to remain constant on time scales of tens of years (Peterson
1993). We thus assume that the flux of the [OIII]$\lambda\,5007$
line or [OI]$\lambda \,6300$ was constant during the monitoring
period. The scaling of spectra was performed by using the method of
\citet{VW92} modified by \citet{Sh04}\footnote{see Appendix A in
\citet{Sh04}}. This method allowed us to obtain a homogeneous set of
spectra with the same wavelength calibration and the same
[OIII]$\lambda$5007 flux.

 In order to investigate the long term spectral variability of AGN
it is necessary to have a consistent  set of  spectra. Since Seyfert
galaxies were observed with different telescopes, in different
position angles, and with different apertures, first we had to
perform corrections for the position angle (PA), seeing and aperture
effects. A detailed discussion on the necessity for these
corrections is given in \citet{Pe93}, and will not be repeated here.

Fluxes of 3C390.3 and NGC 5548 were corrected only for aperture
effects \citep{Sh01, Sh04}, because the host galaxy is weak and the
nucleus have its angle size $<2''$ and  spectra were taken with
apertures $4.2''\times19.8''$, $2.5''\times6''$, and fluxes were
practically independent from seeing and position angle. The fluxes
NGC 4151 were corrected for the position angle (PA), seeing and
aperture effects, because the host galaxy is very bright and
brightness of nucleus is very big and its angle size $>2''$
\citep[see][for details]{Sh08}.

The mean error (uncertainty) for  H$\beta$ and
 the continuum around this line is $<3\%$, while it is $\sim5 \% $ for H$\alpha$.
In order to study the broad components showing the main BLR
characteristics, the narrow components of these lines and the
forbidden lines were removed from the spectra. In this aim, we
constructed spectral templates using the blue and red spectra in the
minimum activity state. The broad and narrow components of
H$\gamma$, He II\,$\lambda$4686, H$\beta$ and H$\alpha$, were fitted
with Gaussians. Then, we scaled the  blue and red spectra according
to our scaling scheme \citep[see the Appendix in][]{Sh01}, using the
template spectrum as a reference. The template spectrum and any
observed spectrum were thus matched in wavelength, reduced to the
same resolution, and  the template spectrum was subtracted from the
observed one. More details about data reduction can be found in
\cite{Sh01,Sh08}.

\begin{sidewaystable}
\caption{BLR characteristics of the three objects.}
\label{tab1}
\centering
\begin{tabular}{llllllll}

\hline
  object &  $R_{H{\beta}}({\rm BLR})$  &                              &Sp               &structure   &Excitation &BLR  \\
(monitor. period) & light           &    $\frac{F_{max}} {F_{min}} $   &  type          &profiles         & model &model\\
             & days            &                                  & (variability)  &                        &              &  \\
          \hline
  NGC 5548    &  6-26                    &5    &1-1.8    &  strong shoulders  &photoion. &turbulent\\
  (1996-2002) & \cite{Pe02}   &     &          &$\pm 1000 \,  {\rm km/s}$    &       &      disk \\
\hline
  NGC 4151  &  1-50            &   6   & 1.5-1.8    & absorption,bumps      &shock&outflow           \\
(1996-2006) &      &       &       & $(-400,-2000)\,  {\rm km/s}$  &photoion.    & possible small disk  \\
\hline
  3C390.3  &  35,100       & 3     &   1-1.8    & double peaked  & photoion.  &disk   \\
(1995-2007) &                   &       &          &   $\pm 4000 \,  {\rm km/s}$ & &  \\
\hline
\hline
\end{tabular}
\end{sidewaystable}

\section{RESULTS}

\subsection{BEL profile variations in NGC 5548}
\label{sec3.1}

This nucleus was observed from 1996 to 2002. A detailed description
of the observations and results can be found in \cite{Sh04}. The
results can be outlined as follows:

-  The flux in the broad lines and in the continuum gradually
decreased, reaching minimum values during May-June 2002. In the
minimum state, the wings of H$\alpha$ and H$\beta$ became extremely
weak, corresponding to a Sy1.8 type and not to a Sy1, as observed
previously when the nucleus was brighter (see Fig.1).

- The H$\alpha$ and H$\beta$ broad line profiles displayed
double-peaked profiles, at radial velocities $\sim \pm 1000 \, {\rm
{km/s}}$ with respect to the narrow components. The relative
intensities and radial velocities  of the peaks varied. During 1996,
the red peak was the brightest, while in 1998 - 2002, the blue peak
became brighter. Their radial velocities vary between $\sim 500$ and
$1200 \, {\rm {km/s}}$ (see Fig. 2).

- In 2000 - 2002, a distinct peak was observed in the red wings of
H$\alpha$ and H$\beta$, at a radial velocity of $\sim +2500 \,
{\rm{km/s}}$. Its radial velocity  decreased from 2000 to 2002 by
 $500 -600 \, {\rm km/s}$
(see Fig. 3).

\begin{figure}
\includegraphics[angle=-90,width=8.5cm,bb=50 60 570 770,clip=]{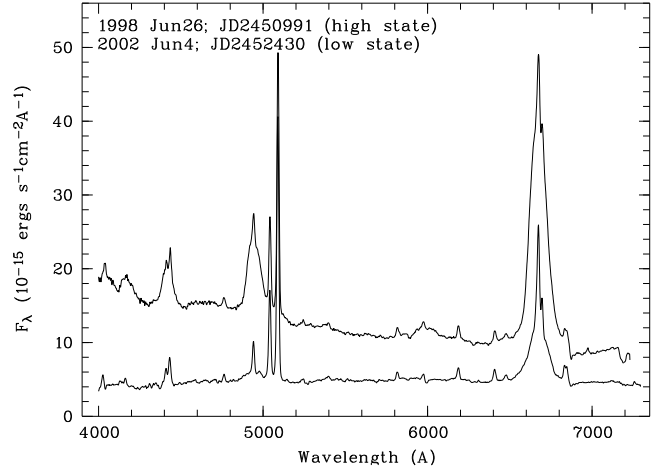}
\caption{The spectra of NGC 5548 corresponding to the high activity
state (top) and the low activity state (bottom).   Observed
wavelengths (we recall that z=0.0167) are displayed on the X-axis,
and fluxes (in units of
$10^{-15}$\,erg\,cm$^{-2}$\,s$^{-1}$\,\AA$^{-1}$) are displayed on
the Y-axis. } \label{fig1}
\end{figure}

\begin{figure}
\centering
\includegraphics[width=8.8cm,bb=84 171 529 690,clip=]{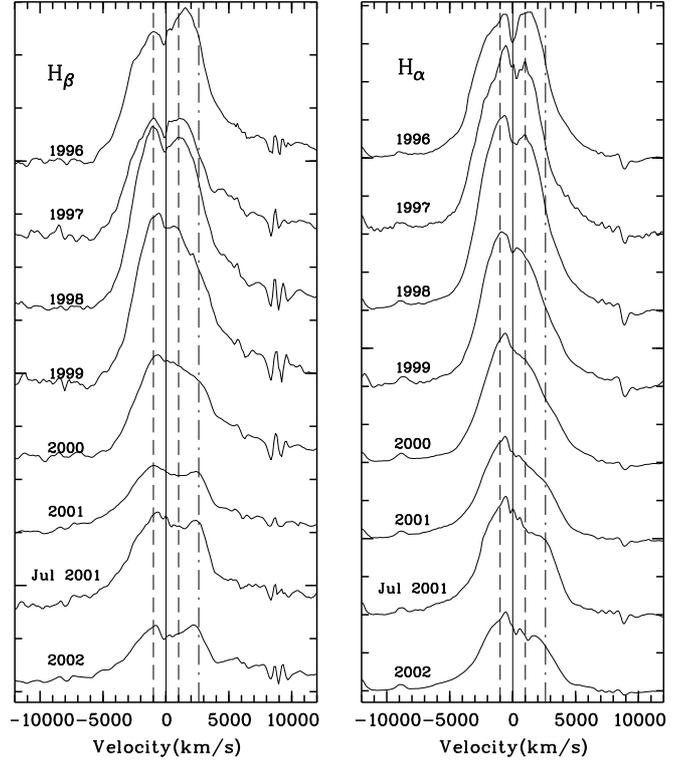}
\caption{ Year averaged observed profiles of the H$\beta$
(respectively the H$\alpha$) broad emission component in NGC 5548,
after subtraction of the continuum and the narrow components $[{\rm
OIII}]\lambda\lambda\,4959,5007$ (respectively $[{\rm
NII}]\lambda6548,6584$ and [{\rm SII}]$\lambda6717,6731$). The
vertical lines correspond to the radial velocities $\pm 1000)\,{\rm
{km/s}}$ - dashed lines, $\sim +2600\,{\rm km/s}$ - dash-doted
lines, $0\,{\rm {km/s}}$ - solid lines. } \label{fig2}
\end{figure}

\begin{figure}
\centering
\includegraphics[width=8.8cm]{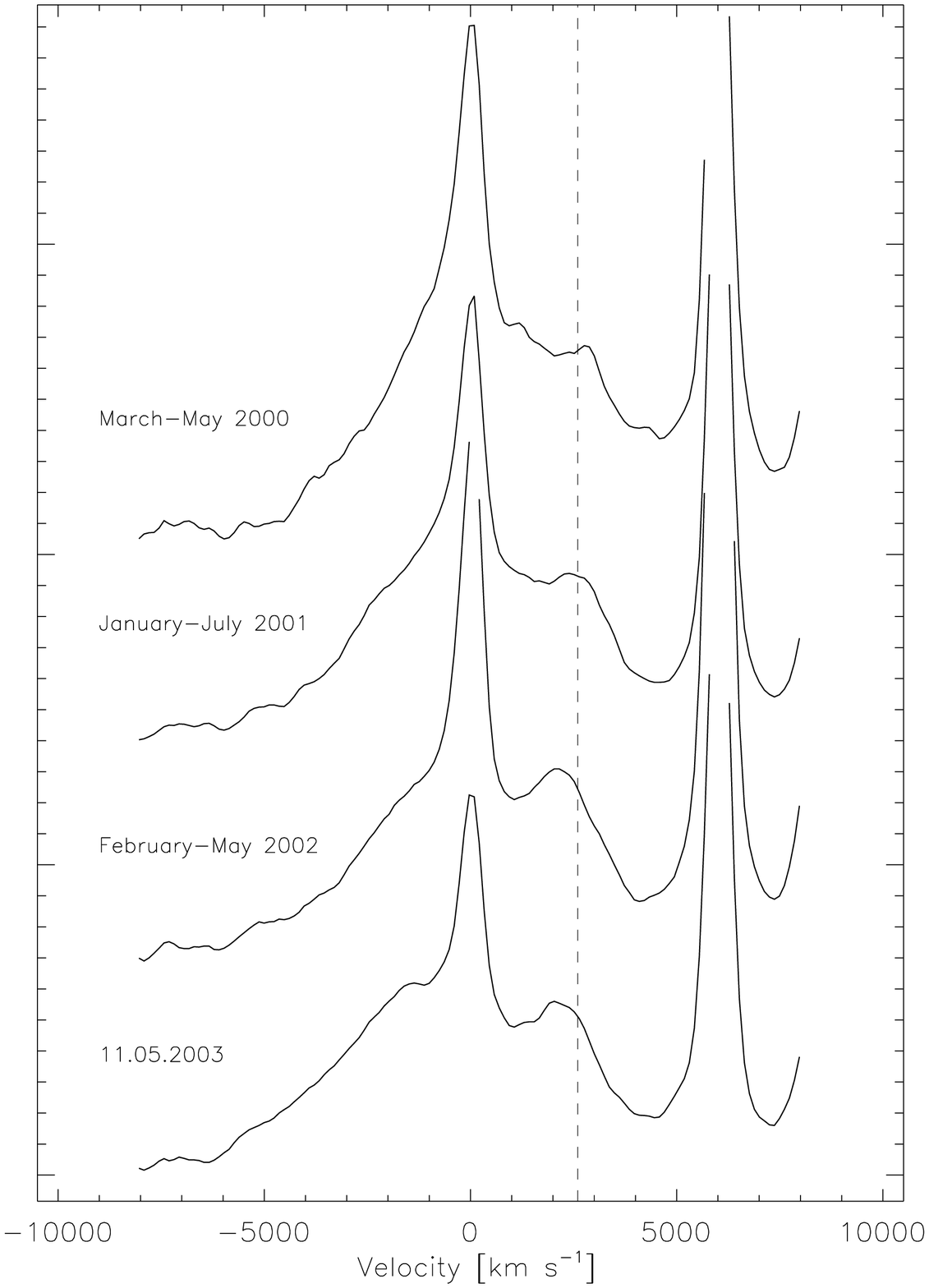}
\caption{ Year-averaged normalized  (with respect to the narrow
H$\beta$ and [OIII] lines) profiles of H$\beta$
 of NGC 5548 in the period 2000--2002.
 The vertical dashed lines correspond to the radial velocity
$\sim +2600\, {\rm km/s}$.} \label{fig3}
\end{figure}

- On small time scales, the flux of different parts of the profiles
changed in almost the same manner, being highly correlated both with
each others (correlation coefficient $r\sim0.94 - 0.98$) and with
the continuum flux  ($r\sim0.88 - 0.97$). These results indicate
that the flux variability on short time scales is caused mainly by
the reverberation effect (see Fig. 4). In Fig. 4 flux 1,2,3,4 and
shape 1,2,3,4 correspond to the flux or shape in the H$\alpha$ and
H$\beta$ profile segments in the following radial velocity
intervals: flux1 and shape1 for ($-3000$, $-2000$)\,km/s; flux2 and
shape2 for ($-1500$, $-500$)\,km/s; flux3 and shape3 for ($+500$,
$+1500$)\,km/s and flux4 and shape4 for ($+2000$, $+3000$)\,km/s.
The shape determination is given in \citet[][3.3.1]{Sh04} using
method suggested by \citet{Wp96}.

 - On long time scales, the variations of  different parts of the line
profiles were mildly correlated with each others and with the
continuum variations, or simply did not correlate at all. So, the
profile changes on long time scales are not due to reverberation, as
earlier discussed  by \cite{Wp96}.

- Changes of the integral Balmer decrement were, on average,
anticorrelated with the continuum flux variations. This is
probably due to an increasing role of collisional excitation as
the ionizing flux decreases. The behavior of the Balmer decrement
in various parts of the line profiles was different in
1996-2000 as compared with 2001.

- From CCF-analysis of the time delay between the continuum
variations and the  H$\beta$ response during 13 years \citep{Pe02},
one finds that the H$\beta$ time lag varied from of 6 days in 2000
to 26 days in 1998-1999.  The relation between the size of the BLR
and the ionizing luminosity is: $R({\rm BLR})\sim L^{0.5}$.

The variation in the broad line profiles of NGC 5548 indicates
presence of an accretion disk emission which contributes to the BEL
fluxes.

\begin{figure*}
\centering
\includegraphics[]{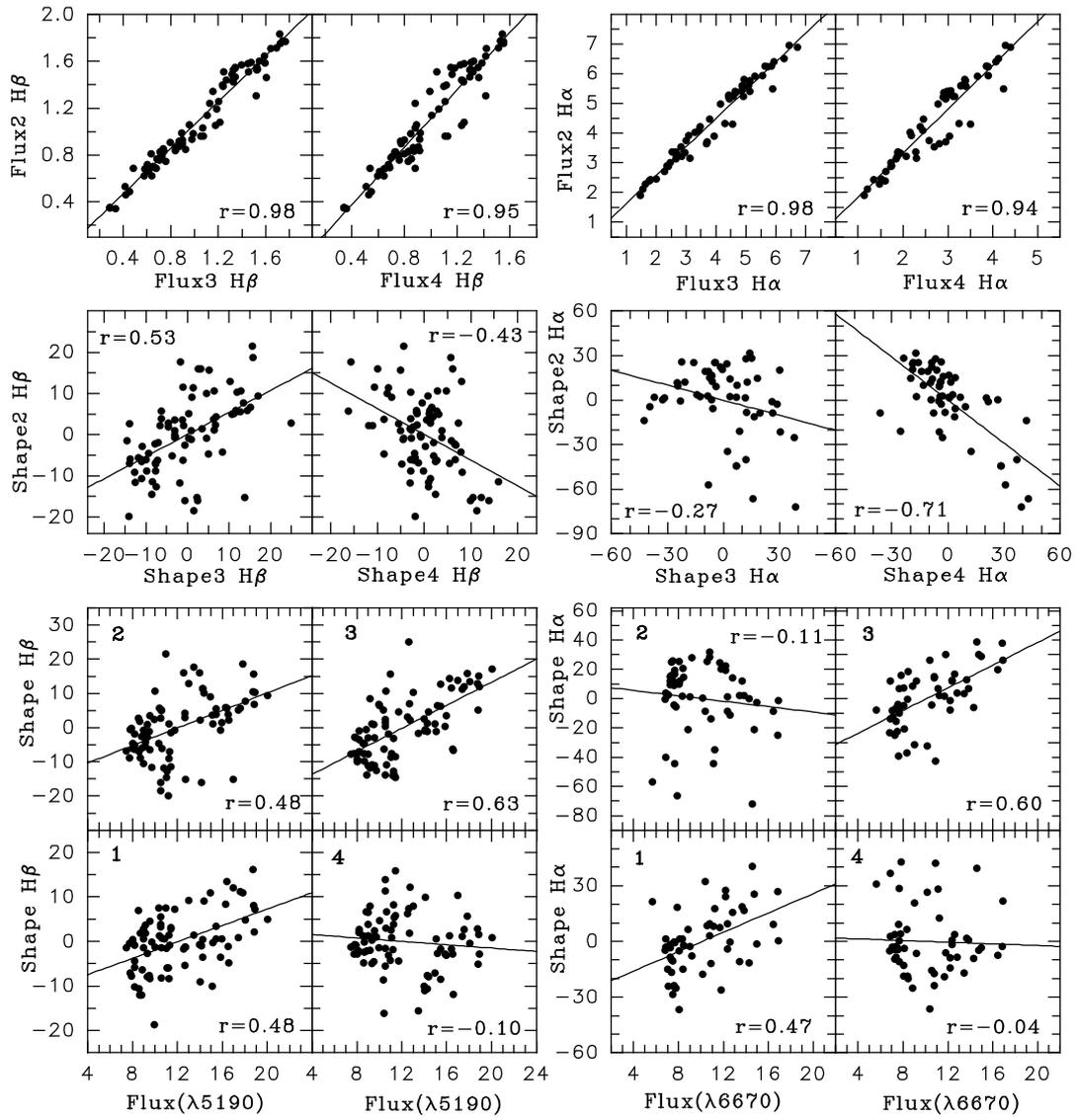}
\caption{  Flux-flux, shape-shape and continuum flux-shape
correlations for different segments of H${\alpha}$ and H$\beta$ in
NGC 5548 (the number of segments is noted inside the plot). The
correlation coefficient r are given inside the plot.} \label{fig4}
\end{figure*}

\subsection{BEL profile variations in NGC 4151}
\label{sec3.2}

The galaxy was observed in period 1996-2006. The details about
observations can be found in \cite{Sh08}. Here we outline some of
the most important results:

- The continuum and line fluxes varied strongly (up to a factor 6)
during  the monitoring period. The emission was maximum in
1996-1998, and there were two minima, in 2001 and in 2005. As a
consequence, the spectral type of the nucleus changed from a Sy1.5
in the maximum activity state to a Sy1.8 in the minimum state (Fig.
5).

\begin{figure}
\centering
\includegraphics[width=8.5cm]{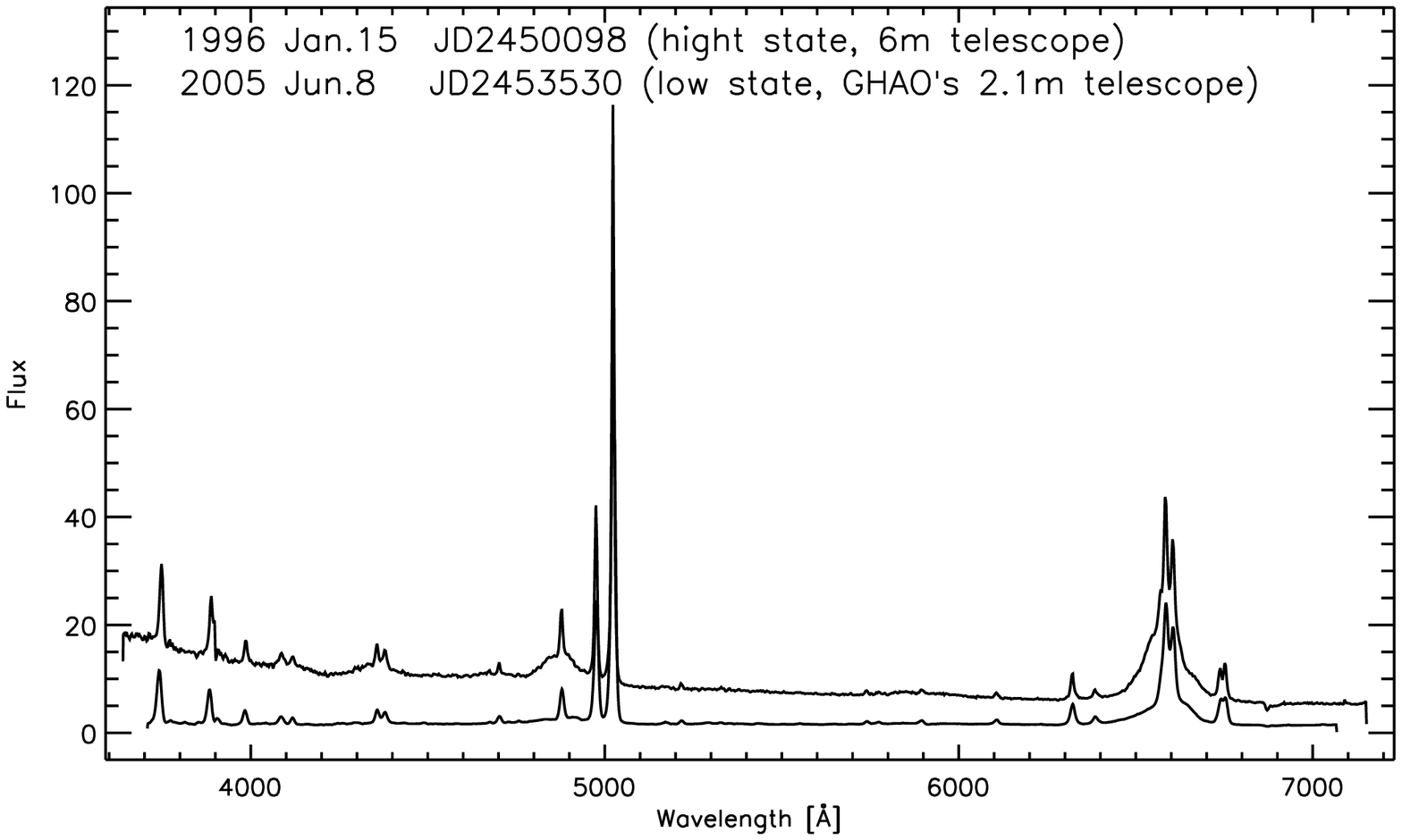}
\caption{The spectra of NGC 4151 corresponding to the high activity
state (top) and to the low activity state (bottom). Observed
wavelengths (we recall that z=0.0033) are displayed on the X-axis,
and fluxes (in units of
$10^{-14}$\,erg\,cm$^{-2}$\,s$^{-1}$\,\AA$^{-1}$) are displayed on
the Y-axis.}\label{fig14151}
\end{figure}

- There are three characteristic periods (1996-1999, 2000-2001 and
2002-2006) in the profile variability. In the first period, when the
lines were the most intense, a highly variable blue component was
observed, which showed two peaks or shoulders at $V_r\sim-4000 \,
{\rm km/s}$ and $V_r\sim-2000 \, {\rm km/s}$ in the rms H$\alpha$
profiles and, to a less extent, in H$\beta$.  In the second period,
the broad lines were much fainter; the feature at $V_r\sim-4000 \,
{\rm km/s}$ disappeared from the blue part of the rms profiles of
both lines;  only the shoulder at $V_r\sim-2000 \, {\rm km/s}$ was
present. A faint shoulder at $V_r\sim3500 \, {\rm  km/s}$ was
present in the red part of rms line profiles. In the third period  a
red feature (a bump or a shoulder) at $V_r\sim2500 \,  {\rm km/s}$
was clearly seen in the red part of both the mean and the rms  line
profiles. The behavior of the rms profiles in the three periods
indicates that the BLR has a complex structure and that its geometry
probably changes with time.

\begin{figure}
\begin{center}
\includegraphics[width=8cm]{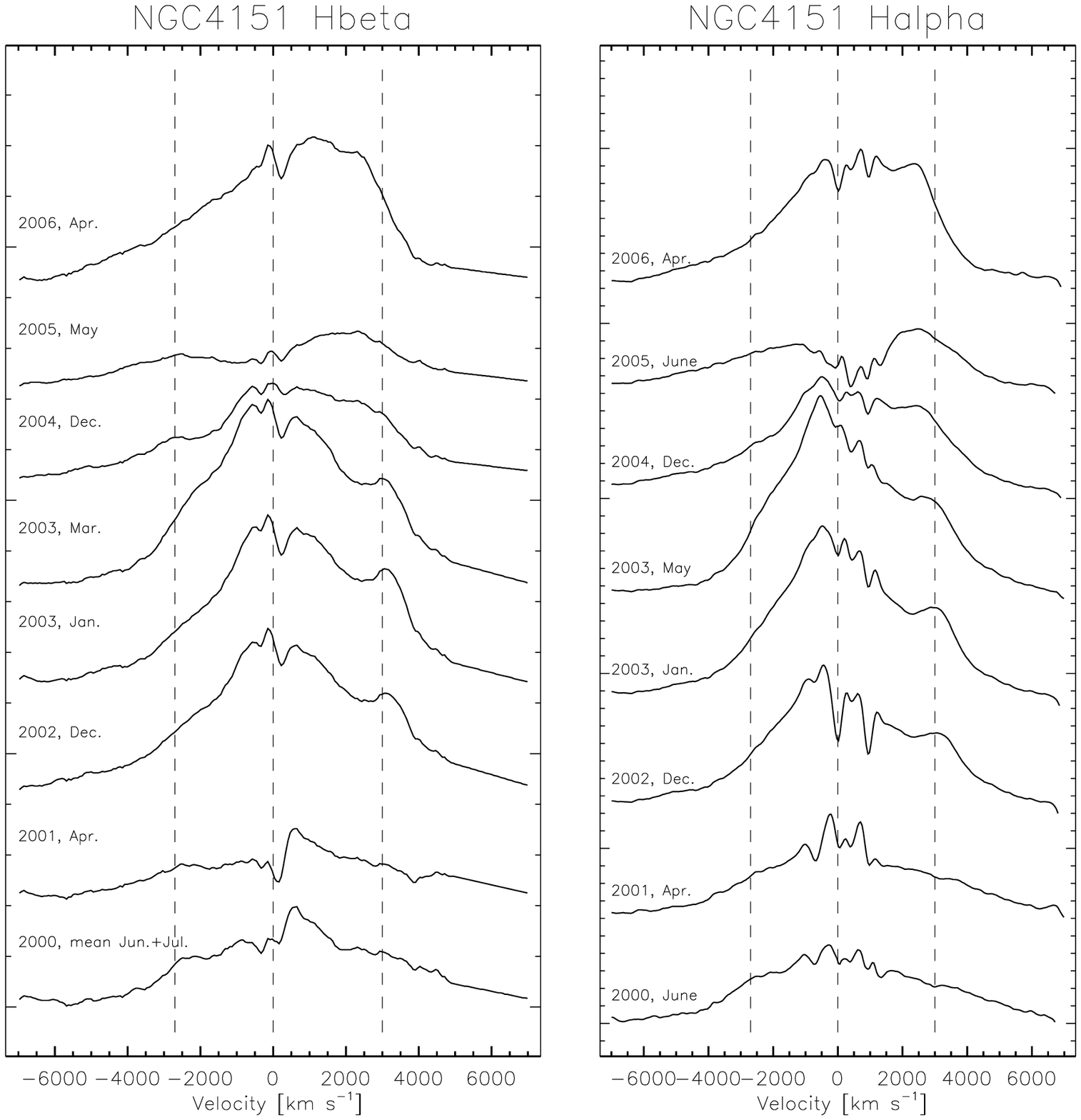}
 \caption{Some examples of month-averaged profiles of the H$\alpha$ and H$\beta$ broad emission
lines from 2000 to 2006 in NGC 4151. The abscissae shows the radial velocities relative
to the narrow components of H$\alpha$ or H$\beta$. The vertical lines correspond
to radial velocities: $-2600 \, {\rm{km/s}}$; $0\,  {\rm km/s}$ and $3000\, {\rm km/s}$. The profiles are shifted
vertically by a constant value.
}
\label{figs}
\end{center}
\end{figure}

\begin{figure*}
\includegraphics[width=5cm]{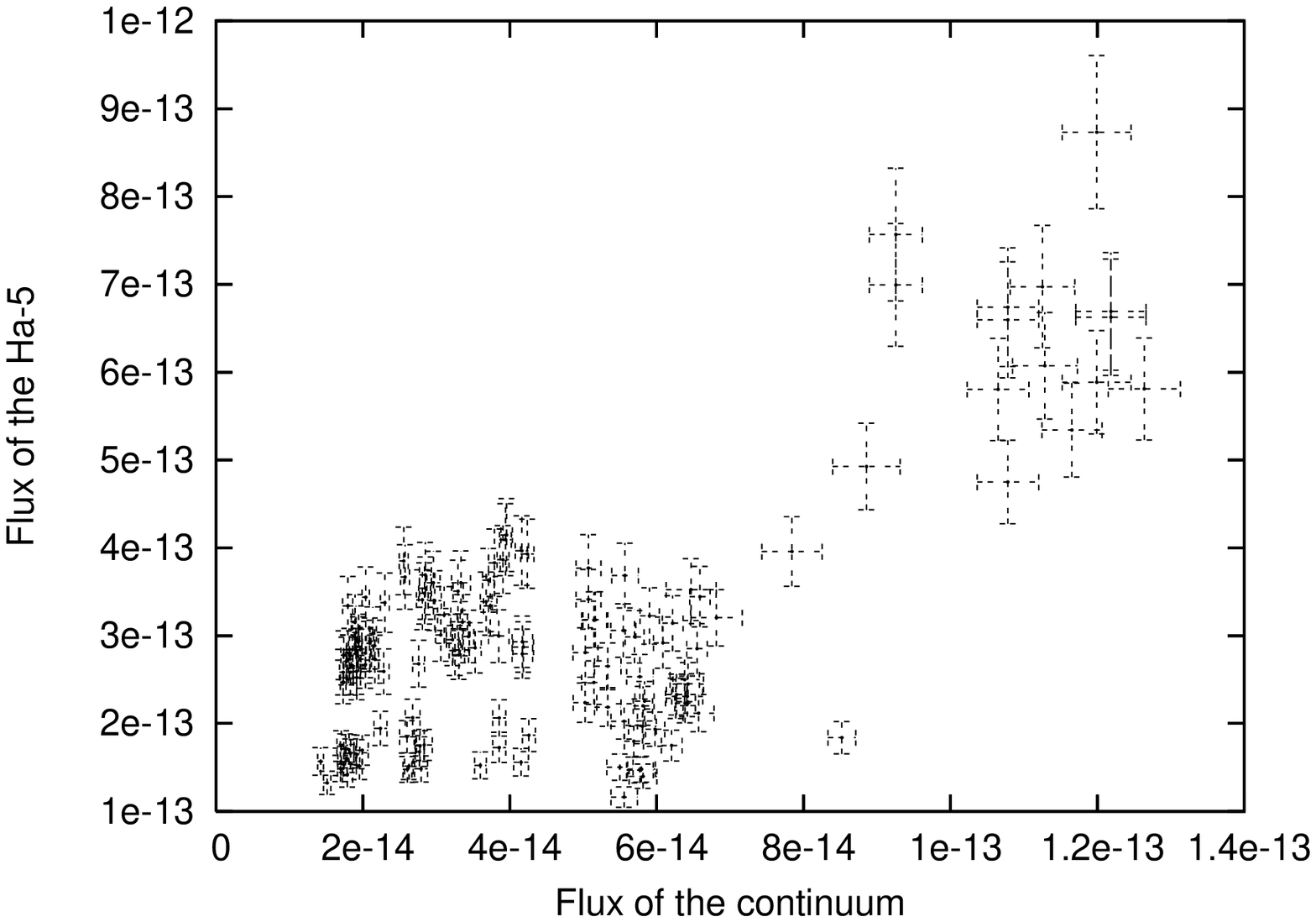}
\includegraphics[width=5cm]{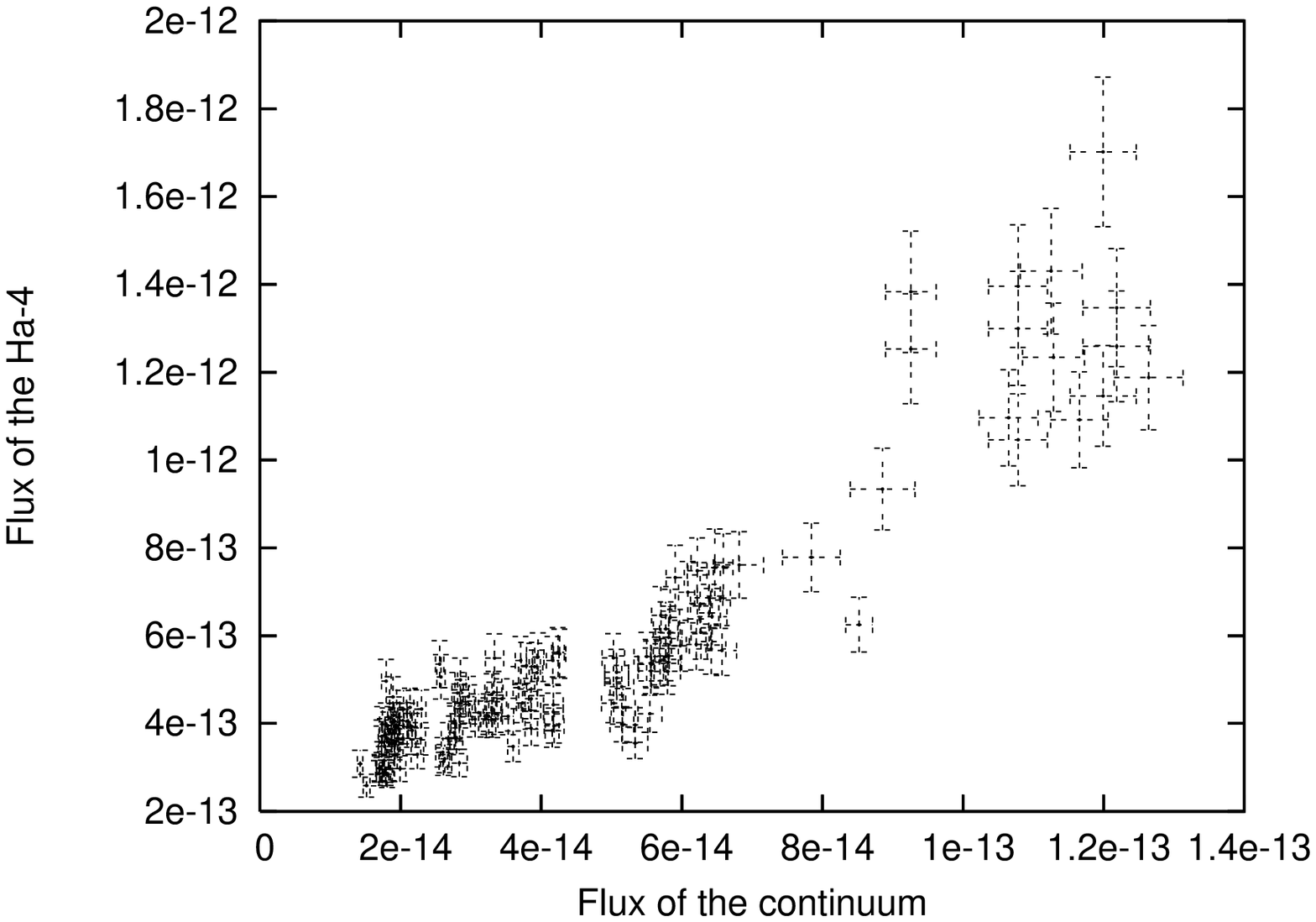}
\includegraphics[width=5cm]{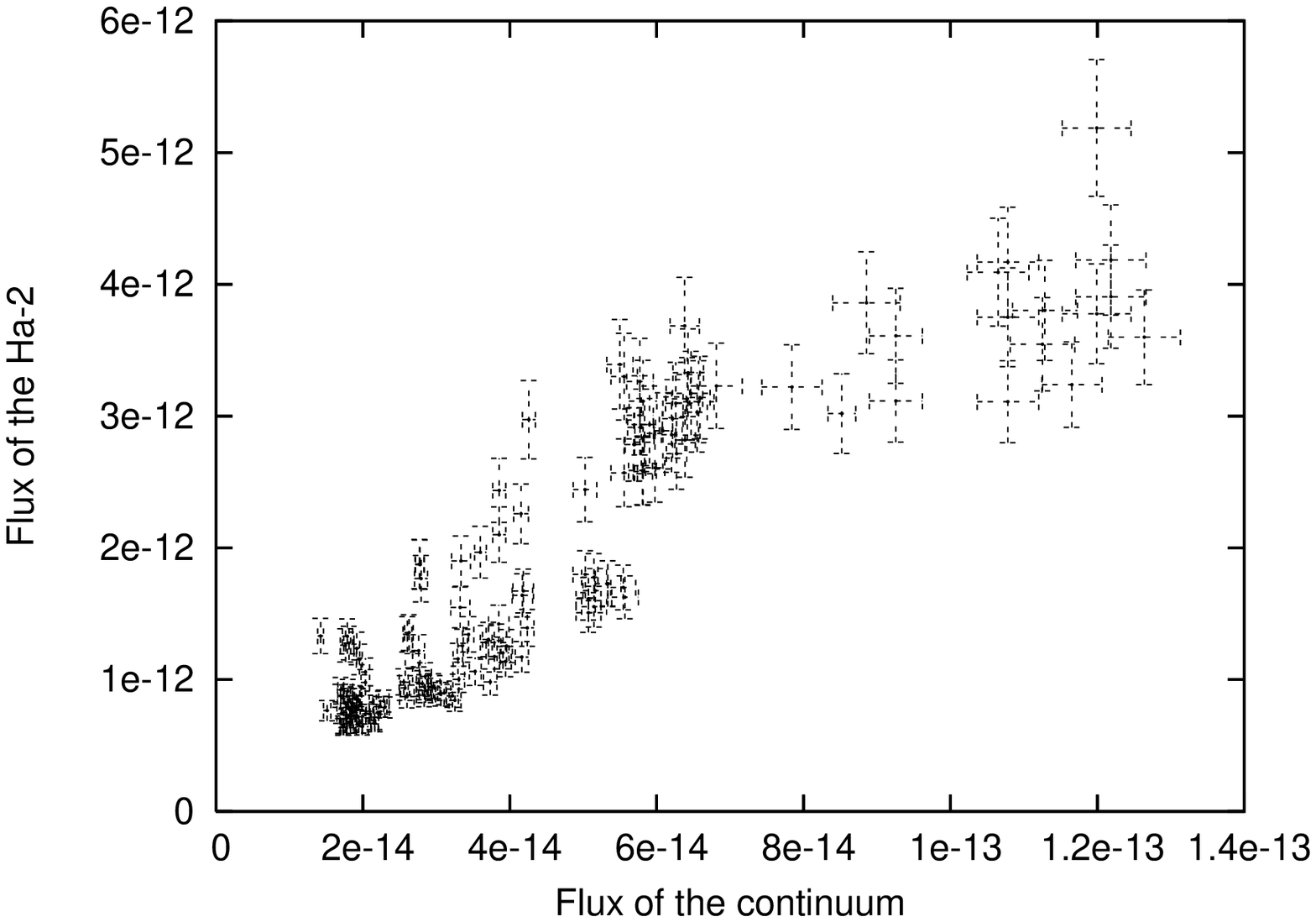}\\
\includegraphics[width=5cm]{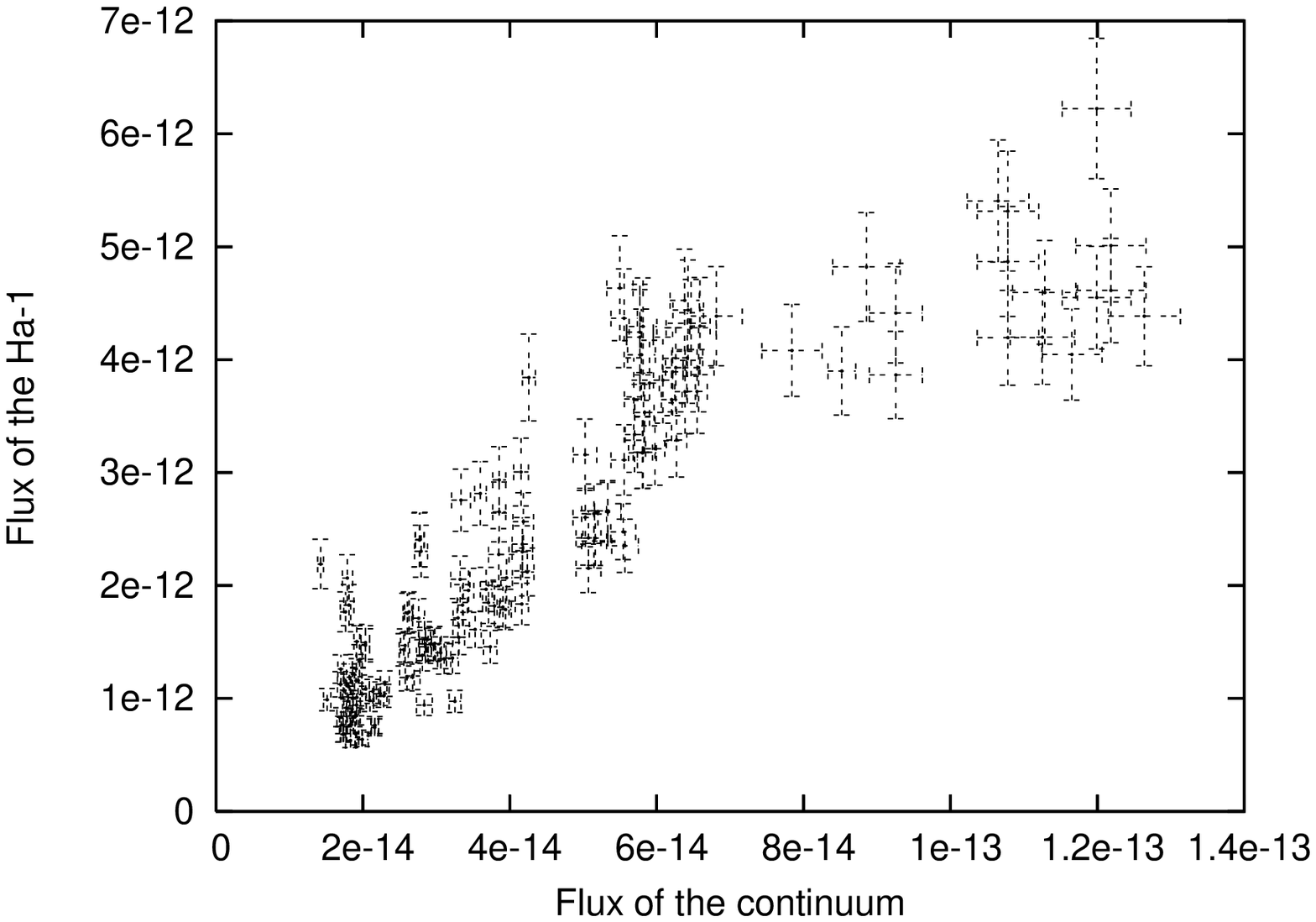}
\includegraphics[width=5cm]{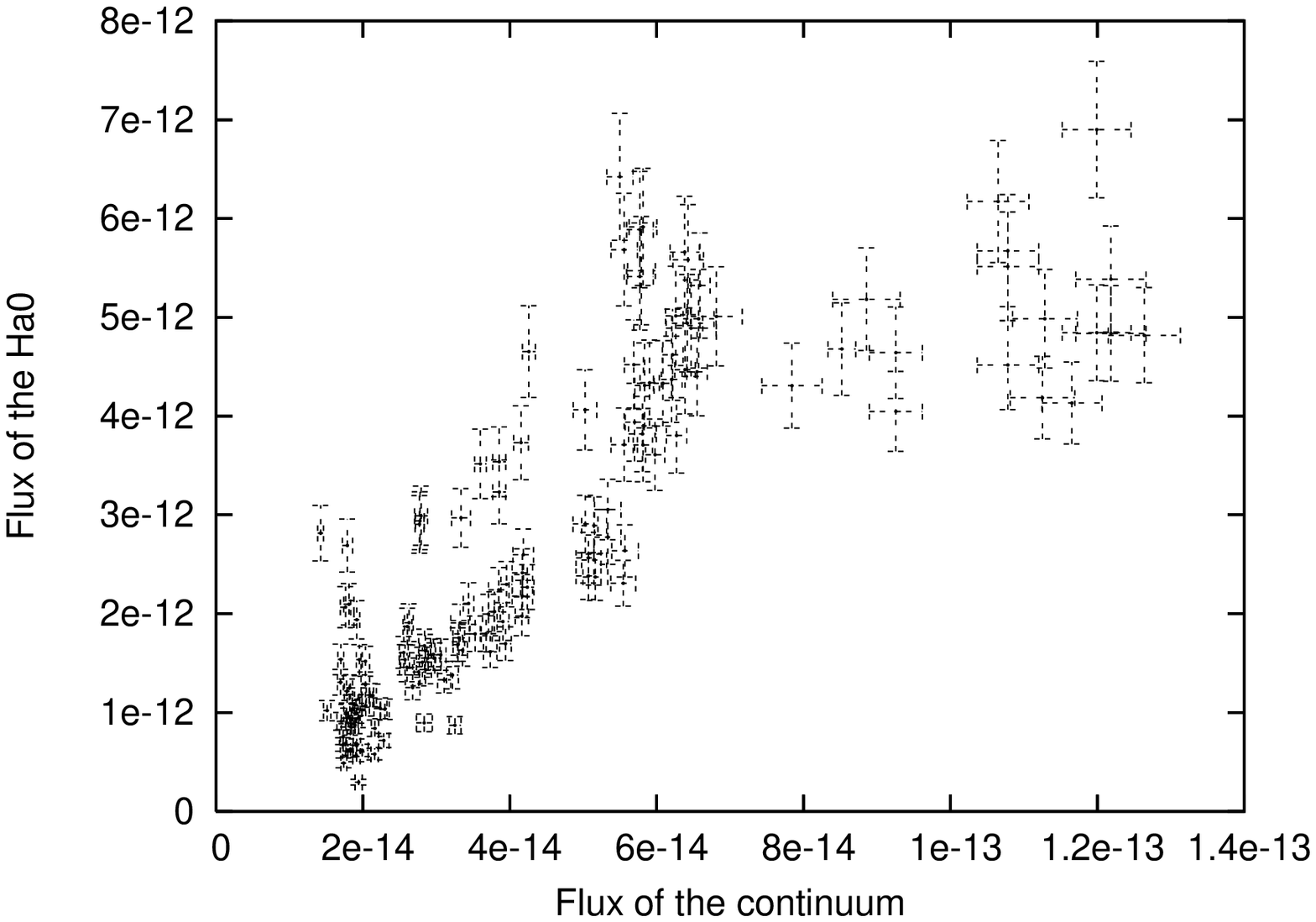}
\includegraphics[width=5cm]{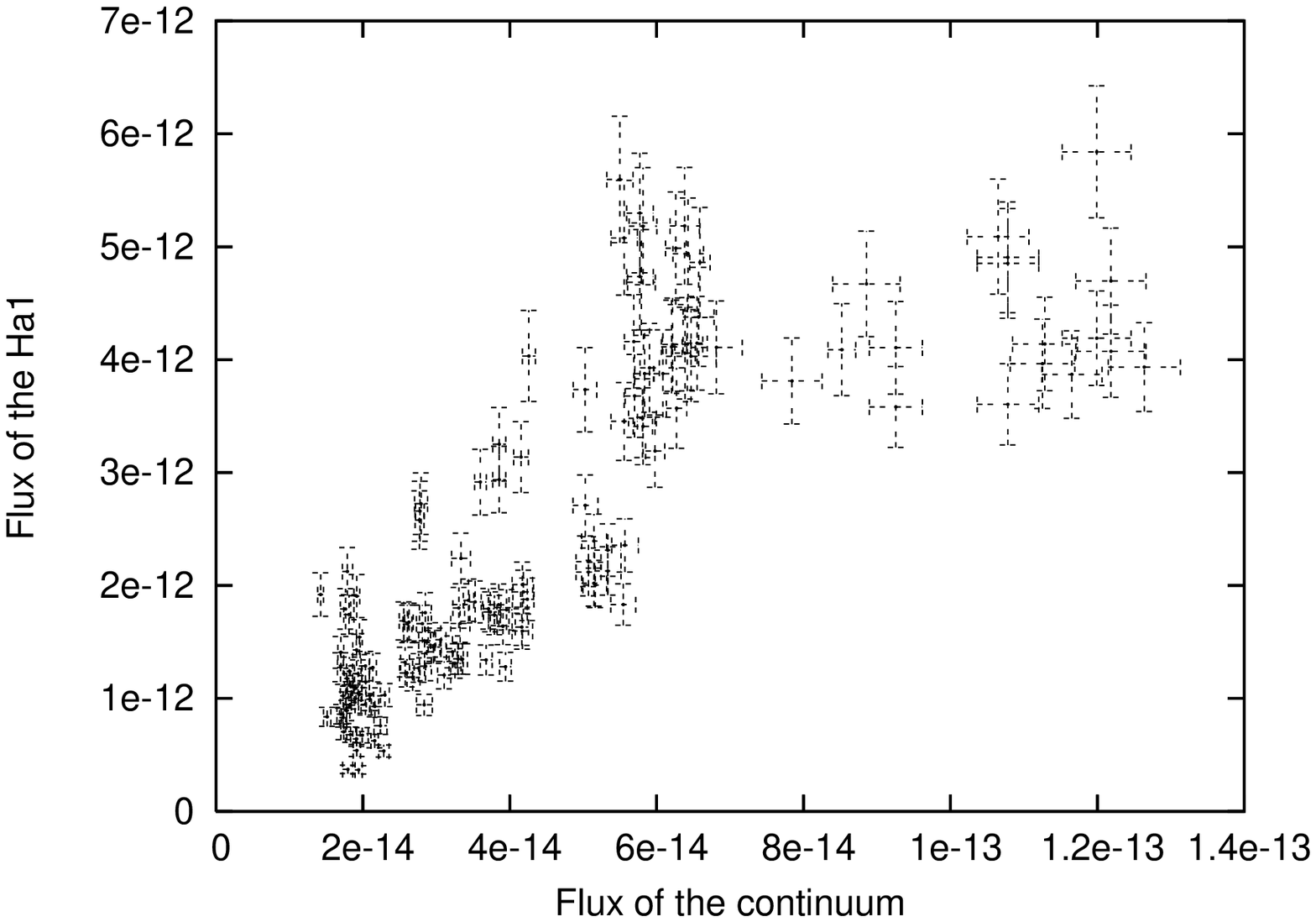}\\
\includegraphics[width=5cm]{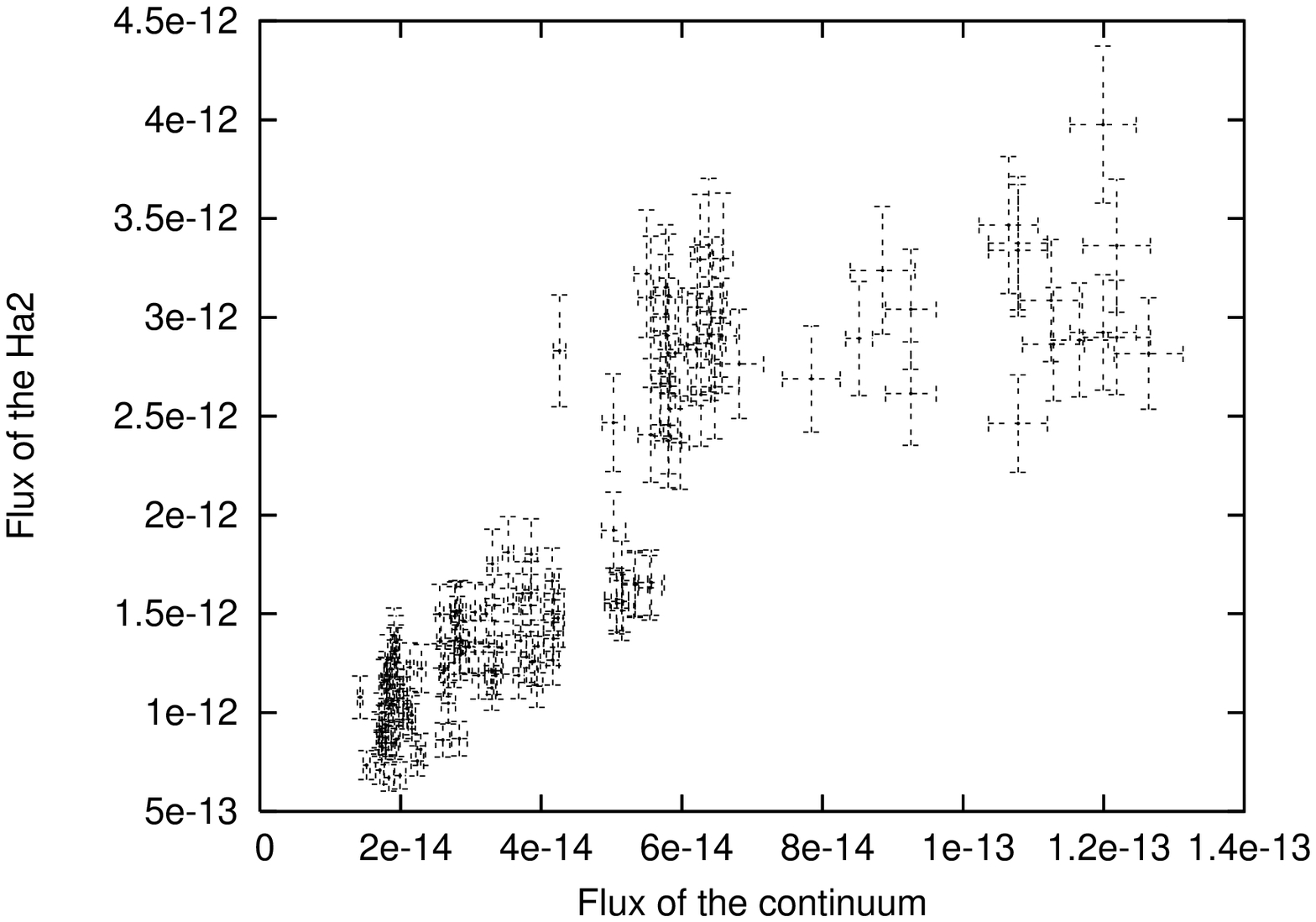}
\includegraphics[width=5cm]{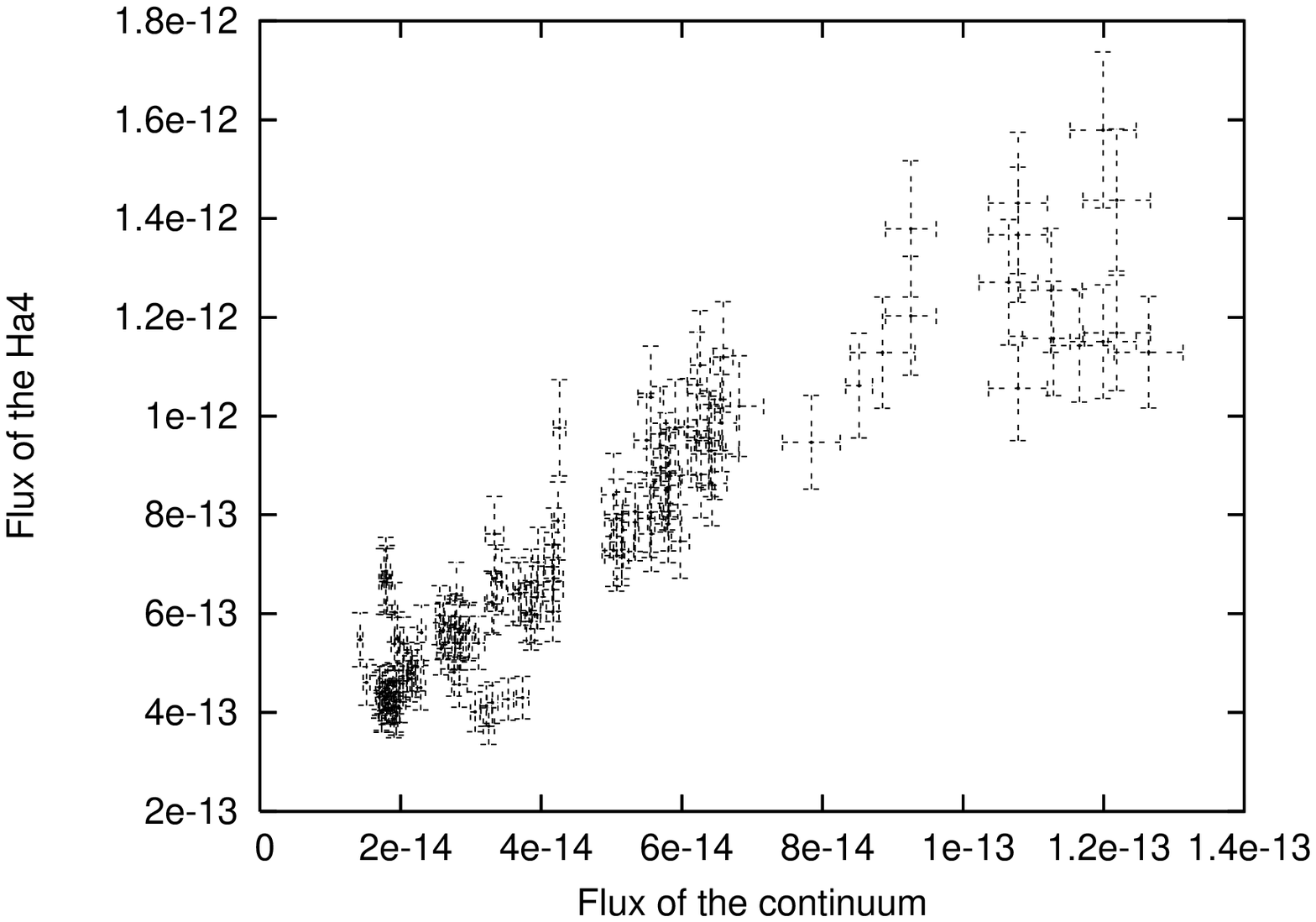}
\includegraphics[width=5cm]{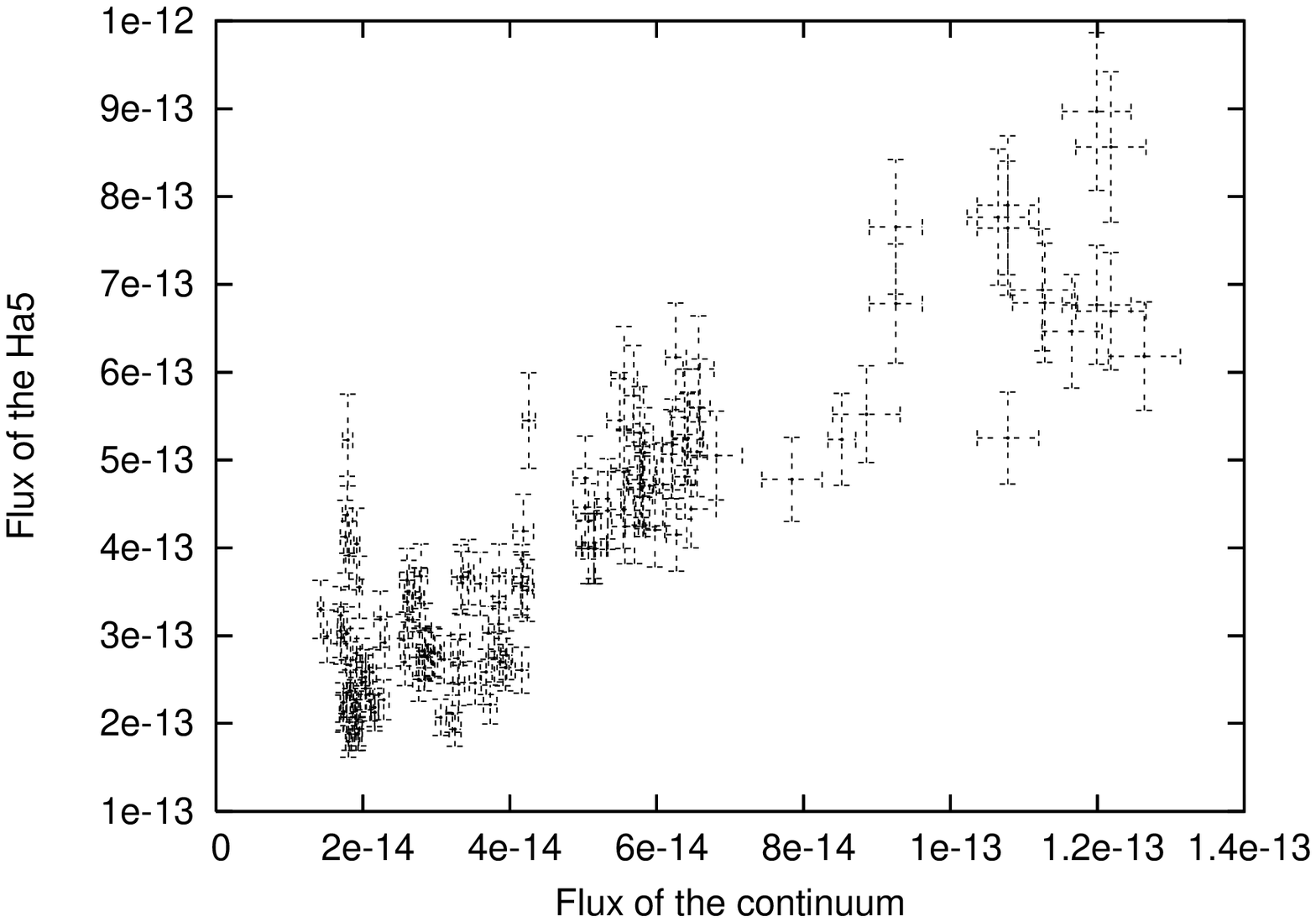}\\
\caption{The flux of different segments of the broad lines as a
function of the continuum flux in NGC 4151. The line flux is given
in $ {\rm erg \ cm^{-2}s^{-1}}$ and the continuum flux in $ {\rm erg
\ cm^{-2}s^{-1}\AA^{-1}}$}. \label{fig9}
\end{figure*}

\begin{figure*}
\includegraphics[width=8cm]{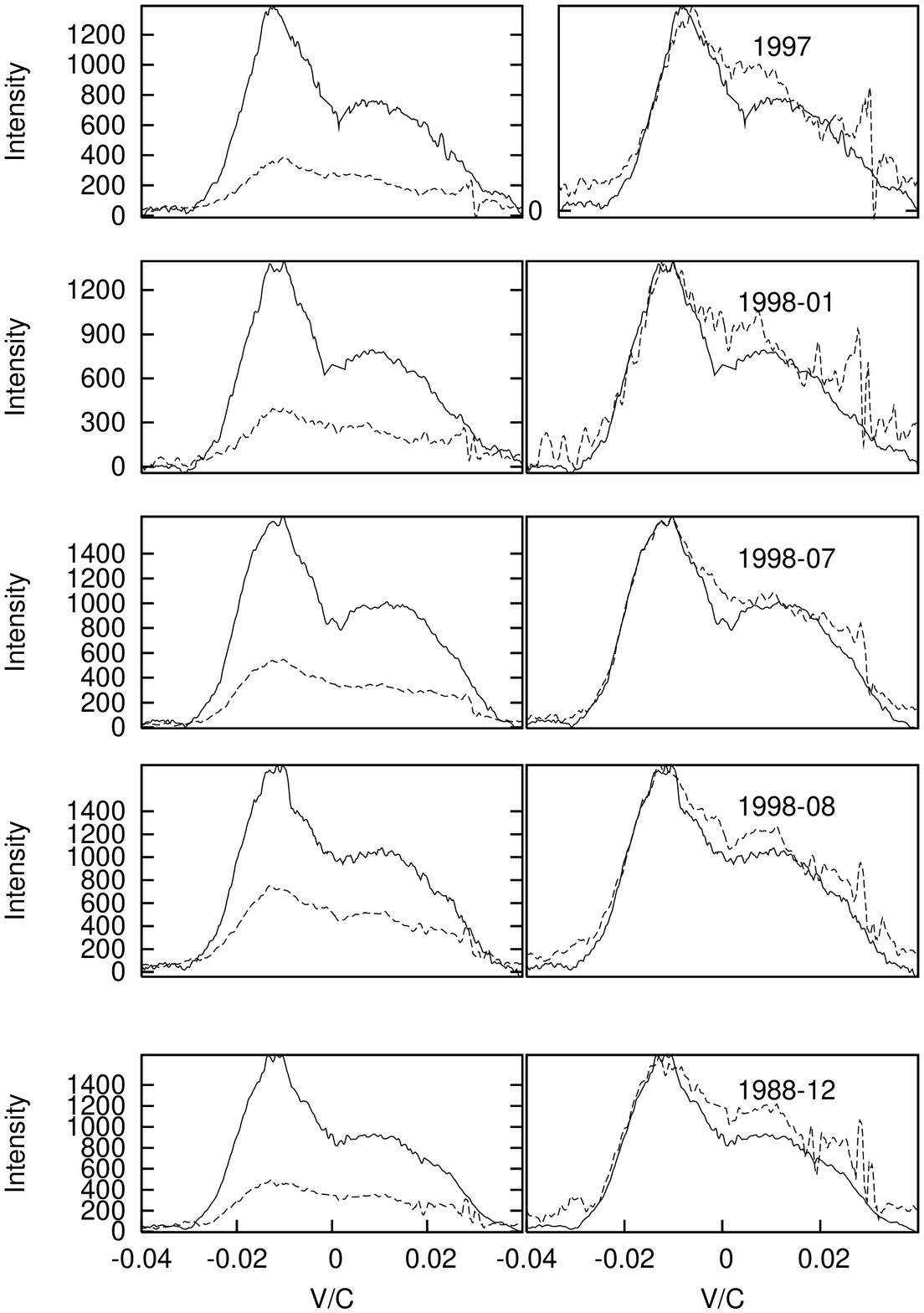}
\includegraphics[width=8cm]{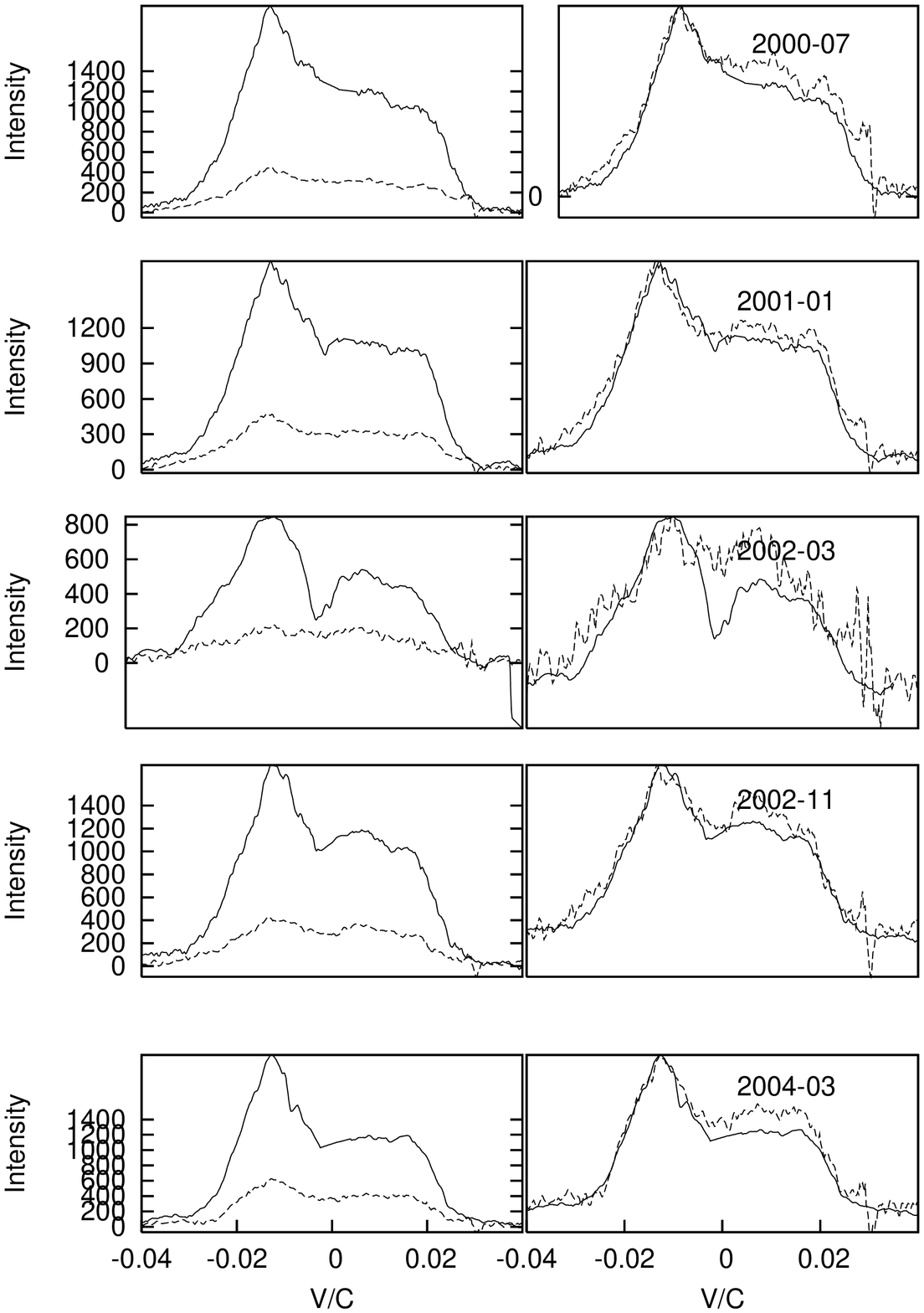}\\
\caption{The profile variations of H$\alpha$ and H$\beta$ in
3C390.3: observed (left side panels) and normalized (right side
panels).  H$\alpha$ and H$\beta$ are denoted with solid and dashed
line, respectively.} \label{ff}
\end{figure*}

In 2005 (May-June), when the nucleus of NGC 4151 was in the minimum
activity state, the line profiles had a structure with two distinct
peaks (bumps) at radial velocities of $(-2600, +2000)\, {\rm km/s}$
in H$\beta$ and $(-1300, +2300) \, {\rm km/s}$ in H$\alpha$. In 2002
a distinct bump appeared in the red wing of both lines at $+3100
\,{\rm km/s}$. The radial velocity of the bump in the red wing
changed from $+3100\, {\rm km/s}$ in 2002 to $\sim2100 \, {\rm
km/s}$ in 2006 (Fig. 6). In 1999-2001 we observed a broad deep
absorption line in H$\beta$ at the radial velocity $V_r\sim-400 \,
{\rm km/s}$.

- The CCFs are often asymmetrical and the time lags between the
lines and the continuum are badly defined, indicating the presence
of a complex BLR, with dimensions of 1 to 50 light-days.

- The Balmer decrement ($BD=\frac{H\alpha} {H\beta} $) was maximum
in $1999-2001$. An anti-correlation with the continuum was observed
in the form of two almost parallel series corresponding to
$1999-2001$ and $2002-2006$  with a difference  in $BD$ by a factor
$\sim 1.5$. In $1996-1998$  the BD did not show any dependence with
the continuum variations. The fact that the BD had the different
values  during the monitoring period (as well as different values
along the profiles) also indicates a multi-- component origin of the
broad lines. It may be caused  either by absorption, or by different
physical conditions in different parts of the BLR \citep{Sh9a}.

- There was a  linear relationship between the emission line and the
continuum flux variations when the continuum flux Fc was small (Fc
$\le 7\cdot 10^{-14} \rm\ erg \ cm^{-2}s^{-1}\AA^{-1}$). The fact
that the H$\alpha$ and H$\beta$ fluxes were well correlated to the
continuum flux means that the ionizing continuum was a good
extrapolation of the optical continuum. When  Fc $\ge 7\cdot
10^{-14} {\rm \ erg\ cm^{-2} s^{-1} \AA^{-1}}$, the line fluxes
either were weakly correlated, or simply did not correlate at all
with the continuum flux. So it seems that the lines were saturated
at high fluxes (see Fig. 7).

\subsection{BEL profile variations in 3C 390.3}
\label{sec3.3}

The observations of 3C 390.3 were performed from 1995 to 2007. The
results of the first part of the monitoring period were published in
\cite{Sh01} and \cite{Sh08}, and those of the last part will be
published soon \citep{Sh9b}. The main results can be outlined as
follows:

-- The broad component of the H$\beta$  and H$\alpha$  lines and the continuum flux
varied by a factor of 3-4  during 1995-2007.

-- The H$\beta$  profiles changed drastically: the wings were very
strong in 1996-1998 (like those of a Sy1), with a prominent bump in
blue wing, at $Vr\sim\, -3700{\rm km/s}$  with respect  to the
narrow H$\beta$ component, and with a less intense  flat bump in the
red wing; in 1997 the wings were very weak (like those of a Sy1.8).
The blue wing was always brighter than the red one in 1995-2007 (see
Fig. 8).

-- The shift of the blue peak in the integral H$\beta$  profiles
closely follows the trend found by \cite{Er97}. These authors have
shown that, if the displacement of the blue bump peak velocity
originates in individual regions associated with a massive binary
BH, then the inferred  rotation period would be about 800 years, and
the corresponding  mass would then be larger than $10^{11}
M_{\odot}$.  They argued that such large binary black hole  masses
are difficult to reconcile with other observations and with theory.
Therefore, they reject the hypothesis of a binary black hole for 3C
390.3. Our results provide further support to their conclusion.

\begin{figure}
\begin{center}
\includegraphics[width=8cm]{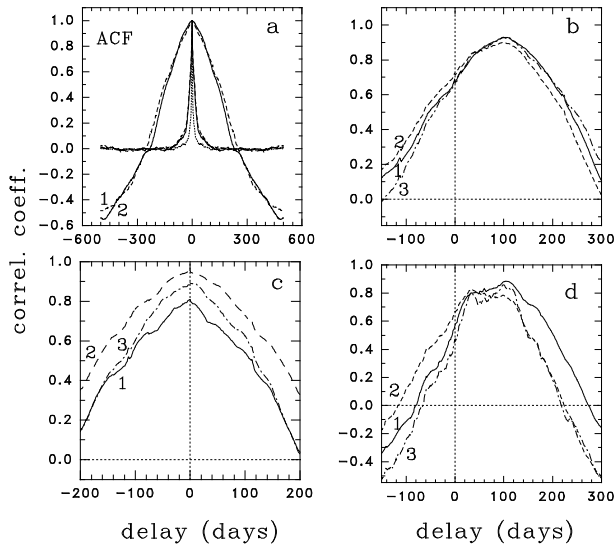}
 \caption{
 The Auto- and Cross-Correlation Functions in 3C390.3. In panel (a), the narrow curve represents the sampling window for the
autocorrelation function (ACFW) between the spectral continuum and
the H$\beta$ line. Also, the autocorrelation functions (ACF) for
H$\beta$  (1) - and the continuum light curves (2) are shown. On
panels (b) and (d), the CCFs for the periods 1995-1999 and 1997-1999
are shown. Curves  (1) represent the CCF between the H$\beta$ flux
and the continuum at 5125 \AA; curves (2) represent the CCF for the
H$\beta$ flux and the continuum, and label (3) refers to the CCF
computed as in case (2), but for a restricted interpolation of 100
days. Panel (c) shows the CCF between: (1) - H$\beta$ blue and red
wings light curves, (2) - H$\beta$ red wing and H$\beta$ core fluxes
and (3) - H$\beta$ blue wing and H$\beta$ core light curves.
 }
\label{fig3c3}
\end{center}
\end{figure}

--  From the cross-correlation analysis, the time lag of the
H$\beta$ emission line response relative to the continuum variations
was found to vary with time: during $1995-1997$ a time lag of about
100 days was detected, whereas for the period  $1998-1999$ a
double-valued time lag of $\sim 100$ days and $\sim 35$ days
(double-peaks in the CCF) is present  in our data (see Fig. 9).

-- The observed flux of the H$\beta$ wings varied
quasi-simultaneously with the same lag relative to the continuum
variations.  Also,  the CCF analysis of the H$\beta$ wings  did not
reveal any delay between the variations of the wings and the central
part of the line, or between two parts of the wings. This behavior
is a strong evidence for predominantly circular motions in the BLR.

-- From the profile differences of H$\beta$, we found that the
variations in velocity of the blue and red bumps and their
differences were anticorrelated  with the flux variations of the
H$\beta$  and with its time lag with respect to the continuum. It
means that the maximum of  the line emissivity  moves across the
disk and corresponds  to smaller radii when  the continuum flux
decreases (the bump velocities increase), and to larger radii when
the continuum flux increases (the bump velocities decrease). These
transient phenomena are expected to result from a variable rate of
accretion close to the black hole. \cite{Sh01}  concluded that these
results favor the formation of the H$\beta$  broad emission line  in
the accretion disk.  The modeling of  the H$\beta$  integral
profiles was performed within the framework of the accretion disk
model, and a satisfactory agreement for the  disk with an
inclination of $25^{\circ}$, and with a region of maximum   emission
located at about $200 R{\rm g}$ was  obtained (R{\rm g} :
gravitation radius).

Recently \cite{St09} fitted the profile of H$\beta$ in 3C390.3
(1995-1999) using a perturbation model \citep{St08} and found that
S--like perturbation in the disk emissivity can explain the observed
line profile variations.

\section{CONCLUSIONS}

Here we briefly outline some results  obtained using long-term
monitoring campaigns of the three AGN. We investigate the variations
of the H$\beta$ and H$\alpha$ line fluxes and profiles, as well as
the correlations between the line and the continuum flux variations.
Our conclusions for the three objects are:

1. For  NGC 5548 our results favor the formation of the broad Balmer
lines in a turbulent  accretion disk containing big moving optically
thick inhomogeneities, able to reprocess the central continuum.

2. The BLR of NGC 4151 seems much more complex. The broad lines are
likely produced at least in two or three  distinct  regions with
different physics (one photoionized by the AGN central source and
another non-radiatively heated).

3. In 3C 390.3 the double-peaked broad emission lines clearly
indicate a disk emission. The variations in line profiles  can be
explained as a perturbation in an accretion disk (more details will
be given in forthcoming paper by \cite{Sh9b}).

In Table 1 we compare some characteristics of variations of the three AGN.
As can be seen, the variability  can be strong (factor 3-6 in the spectral lines),
and during a long period the type of an AGN can change, e.g. from a Sy1 to a Sy1.8.
Two objects show evidence of disk emission.

It is interesting that during a long period, not only the dimension,
but also the geometry of the BLR, can change (as in NGC4151). All
these results should be taken into account when BELs are used  to
estimate the SMBH masses and accretion rates.

 \section*{Acknowledgments}

This work was supported by INTAS (grant N96-0328), RFBR (grants
N97-02-17625 N00-02-16272, N03-02-17123 and 06-02-16843,09-02-01136), State
program 'Astronomy' (Russia), CONACYT research grant 39560-F and
54480 (M\'exico) and the Ministry of Science and Technological
Development of  Serbia through the project
Astrophysical Spectroscopy of Extragalactic Objects (146002).




\begin{thebibliography}{}




\bibitem[\protect\citeauthoryear{Antonucci}{1993}]{An93} Antonucci, R. 1993, ARA\&A, 31, 473

\bibitem[\protect\citeauthoryear{Antonucci \& Miller}{1985}]{AM85}
Antonucci, R. , Miller, J. S. 1985, ApJ, 297, 621

\bibitem[\protect\citeauthoryear{Arshakian et al.}{2008}]{Ar08} Arshakian, T G., León-Tavares, J., Lobanov, A. P., Chavushyan, V. H., Popovi\'c, L.\v C., Shapovalova, A. I., Burenkov, A., Zensus, J. A. 2008, Mem.S.A.It., vol.75

\bibitem[\protect\citeauthoryear{Eracleous et al.}{1997}]{Er97}
Eracleous, M., Halpern, J. P., Gilbert, A. M., Newman, J. A. Filippenko, A. V. 1997, ApJ 490, 216



\bibitem[\protect\citeauthoryear{Osterbrock}{1989}]{Os89}
Osterbrock, D. E. 1989, Astrophysics of gaseous nebulae and active galactic nuclei,
Mill Valley, CA, University Science Books


\bibitem[\protect\citeauthoryear{Peterson}{1993}]{Pe93}Peterson, B.M.\ 1993, PASP, 105, 207

\bibitem[\protect\citeauthoryear{Peterson}{2008}]{Pe08}
Peterson, B. M.  2008, NewAR, 52, 240

\bibitem[\protect\citeauthoryear{Peterson et al.}{2002}]{Pe02}
Peterson, B. M., Berlind, P., Bertram, R. et al. 2002, ApJ, 581, 197

\bibitem[\protect\citeauthoryear{Peterson et al.}{1993}]{Pe93}
Peterson, B. M., Ali, B., Horne, K., Bertram, R., Lame, N. J., Pogge, R. W.,
 Wagner, R. M. 1993, ApJ, 402, 469

\bibitem[\protect\citeauthoryear{Peterson et al.}{1999}]{Pe99}
Peterson, B. M., Barth, A. J., Berlind, P. et al. 1999, ApJ, 510, 659




\bibitem[\protect\citeauthoryear{Shapovalova et al.}{2004}]{Sh04}
Shapovalova, A. I., Doroshenko, V. T., Bochkarev, N. G. et al.
2004, A\&A, 422, 925

\bibitem[\protect\citeauthoryear{Shapovalova et al.}{2001}]{Sh01}
Shapovalova, A.I., Burenkov, A. N., Carrasco, L., Chavushyan, V. H., Doroshenko, V. T., Dumont, A. M. et al. 2001, A\& A, 376, 775

\bibitem[\protect\citeauthoryear{Shapovalova et al.}{2008}]{Sh08}
Shapovalova, A. I., Popovi\'c, L. \v C., Collin, S., et al. 2008,
A\&A, 486, 99S

\bibitem[\protect\citeauthoryear{Shapovalova et al.}{2009a}]{Sh9a}
Shapovalova, A. I., Popovi\'c, L. \v C., Burenkov A. N., et al. 2009a,
A\&A, sent.

\bibitem[\protect\citeauthoryear{Shapovalova et al.}{2009b}]{Sh9b}
Shapovalova, A. I., Popovi\'c, L. \v C., Burenkov A. N., et al. 2009b,
in preparation.

\bibitem[\protect\citeauthoryear{Stalevski et al.}{2008}]{St08}
Stalevski, M., Jovanovi\'c, P, Popovi\'c L. \v C. 2008, Publ. Astron. Obs. Belgrade, 84, 491.

\bibitem[\protect\citeauthoryear{Stalevski et al.}{2009}]{St09}
Stalevski, M., Jovanovi\'c, P, Popovi\'c L. \v C., Shapovalova, A. 2009, Book of Abstracts of VII SCSLSA (eds. L. \v C. Popovi\'c, M.S. Dimitrijevi\'c, D. Jevremovi\'c, D. Ili\'c) p. 58.


\bibitem[\protect\citeauthoryear{Van Groningen \& Wanders}{1992}]{VW92}Van Groningen, E. \& Wanders, I.\ 1992, PASP, 104, 700

\bibitem[\protect\citeauthoryear{Wanders \& Peterson}{1996}]{Wp96}
Wanders, I.,  Peterson, B. M. 1996, ApJ, 466, 174

\bibitem[\protect\citeauthoryear{Weedman \& Khachikyan}{1968}]{W68a}
Weedman, D. W.; Khachikyan, Eh. E. 1968a, Astroph. Zhurnal, 4, 587



\end{thebibliography}
\end{document}